%% file: td.tex
\documentclass[10pt,conference]{IEEEtran}
\IEEEoverridecommandlockouts
\usepackage{cite}
\usepackage{amsmath,amssymb,amsfonts}
\usepackage{algorithmic}
\usepackage{graphicx}
\usepackage{textcomp}
\usepackage{url}
\usepackage[table,usenames, dvipsnames]{xcolor}
\usepackage{balance}
\usepackage{booktabs}
\usepackage{verbatim}
\usepackage{paralist}
\usepackage{subfig}
\usepackage{dirtytalk}
\usepackage[
    bookmarks=false,bookmarksnumbered=true,hypertexnames=false,
    breaklinks=true,linkbordercolor={1 1 1},pdfborder={0 0 0}]{hyperref}
\usepackage{microtype}

\def\BibTeX{{\rm B\kern-.05em{\sc i\kern-.025em b}\kern-.08em
    T\kern-.1667em\lower.7ex\hbox{E}\kern-.125emX}}
\begin{document}

\title{Temporal Discounting in Technical Debt: How do Software Practitioners Discount the Future?}

% ALTERNATIVE AUTHOR STYLE BELOW *******************
%\author{
%\IEEEauthorblockN{Christoph Becker}
%\IEEEauthorblockA{University of Toronto, Canada\\
%{christoph.becker@utoronto.ca}}
%\and
%\IEEEauthorblockN{Fabian Fagerholm}
%\IEEEauthorblockA{University of Helsinki, Finland \\
%Blekinge Inst. of Technology, Sweden\\
%fabian.fagerholm@helsinki.fi}
%\and
%\IEEEauthorblockN{Rahul Mohanani}
%\IEEEauthorblockA{IIIT Delhi, India \\
%rahul.mohanani@iiitd.ac.in}
%\and
%\IEEEauthorblockN{Alexander Chatzigeorgiou}
%\IEEEauthorblockA{University of Macedonia, Greece\\
%{achat@uom.gr}
%}
%}

%% ALTERNATIVE slightly longer option:
\author{
\IEEEauthorblockN{Christoph Becker}
\IEEEauthorblockA{Faculty of Information\\
Univ. of Toronto, Canada\\
{christoph.becker@utoronto.ca}}

\and
\IEEEauthorblockN{Fabian Fagerholm}
\IEEEauthorblockA{Univ. of Helsinki, Finland \\
Blekinge Inst. of Tech., Sweden\\
fabian.fagerholm@helsinki.fi}
\and
\IEEEauthorblockN{Rahul Mohanani}
\IEEEauthorblockA{Department of CSE \& HCD\\
IIIT Delhi, India \\
rahul.mohanani@iiitd.ac.in}
\and
\IEEEauthorblockN{Alexander Chatzigeorgiou}
\IEEEauthorblockA{Dept. of Applied Informatics \\
Univ. of Macedonia, Greece\\
{achat@uom.gr}
}
}

% ************************************************************
% ****************** ALTERNATIVE longer option: **************
% ************************************************************

%\author{
%\IEEEauthorblockN{Christoph Becker}
%\IEEEauthorblockA{Faculty of Information\\
%University of Toronto, Canada\\
%{christoph.becker@utoronto.ca}}

%\and
%\IEEEauthorblockN{Fabian Fagerholm}
%\IEEEauthorblockA{Department of Computer Science, University of Helsinki, Finland \\
%Blekinge Institute of Technology, Sweden\\
%fabian.fagerholm@helsinki.fi}
%\and
%\IEEEauthorblockN{Rahul Mohanani}
%\IEEEauthorblockA{Department of CSE \& HCD\\
%Indraprastha Institute of Information Technology Delhi, India \\
%rahul.mohanani@iiitd.ac.in}
%\and
%\IEEEauthorblockN{Alexandros Chatzigeorgiou}
%\IEEEauthorblockA{Department of Applied Informatics \\
%University of Macedonia, Thessaloniki, Greece\\
%{achat@uom.gr}
%}
%}

\maketitle

\begin{abstract}

Technical Debt management decisions always imply a trade-off among outcomes at different points in time. In such intertemporal choices, distant outcomes are often valued lower than close ones, a phenomenon known as temporal discounting. Technical Debt research largely develops prescriptive approaches for how software engineers should make such decisions. Few have studied how they actually make them. This leaves open central questions about how software practitioners make decisions.

This paper investigates how software practitioners discount uncertain future outcomes and whether they exhibit temporal discounting. We adopt experimental methods from intertemporal choice, an active area of research. We administered an online questionnaire to 33 developers from two companies in which we presented choices between developing a feature and making a longer-term investment in architecture. The results show wide-spread temporal discounting with notable differences in individual behavior. The results are consistent with similar studies in consumer behavior and raise a number of questions about the causal factors that influence temporal discounting in software engineering. As the first empirical study on intertemporal choice in SE, the paper establishes an empirical basis for understanding how software developers approach intertemporal choice and provides a blueprint for future studies.

\end{abstract}


\section{Introduction}
%\textcolor{red}{I believe we must reduce the Introduction  a bit, and definitely provide the main RQ in this section, and the main motivation just preceding the RQ.}

Many Technical Debt (TD) management decisions are intertemporal choices: They involve a trade-off among uncertain outcomes at different points in the future. This represents the intertemporal nature of trade-offs implied in the concept of technical debt as articulated by McConnell, who located the source of technical debt in ``decisions that are `expedient' in the short term but \ldots{} costly in the long term'' \cite{mcconnell_technical_2007}, following Cunningham’s original article \cite{cunningham_wycash_1992}. The temporal nature of technical debt remains a key element of the commonly accepted definition that ``technical debt is a collection of design or implementation constructs that are expedient in the short term, but set up a technical context that can make future changes more costly or impossible'' \cite{avgeriou_managing_2016}. While technical debt is understood to be incurred often unintentionally, TD management needs to address questions such as---\textit{When to fix a code smell?}, \textit{Is it worth to invest effort now for refactoring?} and \textit{Should this particular element of technical debt be paid back now, at the next release, later, or never?}  Such intertemporal decisions involve actors in the systems development process who make choices that often involve a trade-off between uncertain future outcomes. 

In response to these questions, TD research largely develops prescriptive approaches for how software developers and architects should make TD management decisions. Of the 240 papers involving trade-off decisions studied in a recent literature review, 37\% proposed, and 27\% evaluated, a proposed solution \cite{becker_trade-off_2018}. Solution proposals range from sophisticated financial modeling techniques to optimize the TD portfolio to software quality metrics and tools that are meant to inform the decision making process by providing measurable statistics and indicators to feed into the decision making process. 

The degree to which individuals discount future outcomes is represented by temporal discounting concepts such as discount rates. In prescriptive models, future outcomes are often discounted using a fixed financial interest rate in order to establish the supposed present value of a future outcome. Descriptive intertemporal choice research in disciplines including psychology, behavioral economics and management science has firmly established, however, that people facing actual intertemporal choices will discount future outcomes at rates that are non-linear and may change over time.

While TD decisions always imply and often explicitly involve temporal discounting, few have studied what really happens in that process, i.e., how software practitioners actually make those decisions. Only nine of the 240 TD papers examined in a recent review \cite{becker_trade-off_2018} performed empirical studies of concrete TD decisions made in software projects. Even those nine studies were found not to investigate temporal decision making using focused methods and theories. This corresponds to a general neglect of descriptive research on decision making with intertemporal parameters in software engineering research \cite{becker_intertemporal_2017}. 

Many important questions about how software engineers make decisions remain unexplored. This study begins to address a question central to Technical Debt:

\smallskip
{\narrower \noindent \textit{How do software practitioners discount uncertain future outcomes?}\par}
\smallskip

This overarching question guides our research design. We describe a study that explores the implied discount rates of software developers when faced with TD choices. We discuss the findings in the light of the extensive literature on intertemporal choice in psychology and behavioral economics. 

\section{Background}
\subsection{Intertemporal Choice}

TD management is full of intertemporal choices: ``decisions involving trade-offs among costs and benefits occurring at different times'' \cite{frederick_time_2002}. These trade-offs appear explicitly and implicitly in TD: Implicitly, temporal discounting influences behaviors that disregard temporally distant outcomes without a conscious choice, which is often regarded as a source of TD. Explicitly, many TD management decisions need to address the temporal trade-off directly. 

Intertemporal choice is a very active area of research in psychology, behavioral economics, neuro-economics, management science, marketing, and other fields \cite{frederick_time_2002,soman_psychology_2005,loewenstein_time_2003,loewenstein_neuroeconomics_2007}. Intertemporal choice research aims to explain from neurological and psychological perspectives how the mind makes choices when faced with intertemporal decisions, understand under which conditions people discount the future disproportionately, model and predict consumer behavior, and understand how the architecture of choice influences outcomes. Despite decades of research, no undisputed consensus has emerged over the precise mechanisms that influence temporal discounting and the best theoretical models to describe and explain it \cite{frederick_time_2002}. Nevertheless, this body of research provides powerful theories, research methods, experiment designs, empirical guidelines, and conceptual frameworks that support researchers in exploring, understanding, modelling and predicting how people will make intertemporal choices \cite{loewenstein_time_2003,loewenstein_neuroeconomics_2007,soman_psychology_2005,weber_experience-based_2006}.

The intertemporal choice behavior \textit{of consumers} in particular has been explored in depth in hundreds of studies\cite{frederick_time_2002}. A typical and frequently administered basic experiment involves an explicit trade-off between a monetary reward at a certain point in time and a (higher) monetary reward at a later point in time. By establishing how much more the average consumer expects to receive for the later reward to be equally valuable as the closer reward, these experiments establish a discount rate that describes, in numerical terms, how experiment participants discount future outcomes. 

\begin{table*}    
\centering
\caption{Selected experiments in intertemporal choice. For reviews, see \cite{frederick_time_2002,soman_psychology_2005}.}
    \label{tab:experiments}
    \begin{tabular}{|p{1.3cm}|p{4cm}|p{4.6cm}|p{1.7cm}|p{4.2cm}|}
    \toprule
\textbf{Citation} & \textbf{Sample question} & \textbf{Elicitation method and task description} &  \textbf{Sample} & \textbf{Effects} \\\midrule
        Thaler 1981 \cite{thaler_empirical_1981} & ``\dots you won some money in a lottery. You could take the money now or wait until later. How much would you require to make waiting just as attractive as getting the money now?'' & \textit{Matching:} A 3x3 table to be completed. For each time frame (3m, 1 yr, 3yrs) and prize amount (varied from \$15-3000), the participants fill in a matching amount. The matching amount provides the indifference point.  & 20 usable responses from students & Implicit discount rates are very large; discount rates drop sharply as the prize or the time increases.\\\hline
        Coller \& Williams 1999 \cite{coller_eliciting_1999} & ``\dots you have a choice of two payment options; option A or Option B. If you choose B you will receive a sum of money 3 months from today. If you choose A you will receive a sum of money 1 month from today.'' & \textit{Choice}: A multiple price list presents 15 alternatives, with either 500\$ in a month, or 500\$+x in 3 months (where x is used to calculate the individual discount rate IDR) The point at which preferences switch between now and then is used to calculate the indifference point. & 199 graduate and undergraduate students & Mean Annual Interest Rate of $17.5\%-20\%$. Providing information on the AR associated with future payments (in \%) or the available market rates reduces individual discount rates significantly.  \\\hline
        Harrison et al 2002 \cite{harrison_estimating_2002} & ``You will have a choice of two payment options; Option A or Option B. If you choose Option B you will receive a sum of money 7 months from today. If you choose Option A, you will receive a sum of money 1 month from today, but this Option (A) will pay a smaller amount than Option B.'' &
        \textit{Choice}: Subjects were given payoff tables as in Coller \& Williams. The basic question was modified in six ways: E.g. successive questions with increasing x, simultaneously posing several questions with varying x, providing two future income options, four possible time horizons (6m, 12m, 24m, 36m), etc. &  268 people & Discount rates are constant over the one- to three-year horizons. Discount rates vary significantly with respect to socio-demographic variables.\\\hline
        Zauberman et al 2009 \cite{zauberman_discounting_2009} & ``Multiple experiments explore a set of related factors that link the subjective perception of time to the observable behavior of temporal discounting. In Experiment 1, subjects were asked to put a ‘value’ on delaying an outcome (gift certificate). Subjective assessments of differing prospective time horizons were compared to the changes in objective time horizons.'' & \textit{Matching:} Participants first were presented with a scenario to imagine receiving a gift certificate worth \$75. They were then told that the gift certificate was valid that day and were asked to indicate how much they would have to be paid in order to wait for 1 month (1 year or 3 years) before using the gift certificate. Participants were also asked to mark the perceived length of that time frame visually on a graphical scale ranging from `very short' to `very long'. &  57~un\-der\-gradu\-ate students (experiment 1), 106 students (experiment 2) & The standard pattern of hyperbolic discounting is explained with subjective time perception: Subjective estimates of future time horizon change less than the corresponding change in objective time. Rates of discounting decrease with increased time intervals not because people's internal discount functions are approximated by hyperbolas, but because discount rates are calculated using objective time horizons. Internal discount rates calculated using subjective estimates of time horizon do not decrease over time. \\ \bottomrule
    \end{tabular}
\end{table*}

\subsection{Temporal Discounting}
While some individuals in some situations may exhibit negative discount rates, the expected behavior normally involves positive discount rates. This is normal, since the present simply looms larger than the future. Laboratory experiments of temporal discounting effectively predict real-world behavior in many domains, including ``credit card debt, smoking, exercise, body-mass index, and infidelity''~\cite[p.3]{hardisty_how_2011}. Temporal discounting on its own does not explain \textit{why} individuals may value outcomes differently depending on their timing: Many factors can influence a person’s temporal discounting. Frederick et al.\cite{frederick_time_2002} clarify their terminology as follows: 

\begin{quote}
We distinguish time discounting from time preference. We use the term \textit{time discounting} broadly to encompass any reason for caring less about a future consequence, including factors that diminish the expected utility generated by a future consequence, such as uncertainty or changing tastes. We use the term \textit{time preference} to refer, more specifically, to the preference for immediate utility over delayed utility.
\end{quote}

In this paper, we focus on taking the first step of intertemporal choice research in TD management: We aim to empirically establish \textit{whether} software practitioners exhibit temporal discounting. On that basis, we will begin to explore the causal factors that can explain observed behavior and discuss several theoretical and empirical perspectives on \textit{how and why} this behavior comes about.

A comprehensive and widely cited review of intertemporal choice literature summarizes dozens of studies \cite{frederick_time_2002}. Empirical studies use a range of methods, but often present participants with a choice or a matching task. In a \textit{choice} task, participants choose the preferred outcome out of two options; in a \textit{matching} task, they `fill in a blank' to indicate, for example, the reward or price that would make one outcome equivalent to a given outcome at a different point in time. The choice task provides upper or lower bounds on the \textit{indifference point} at which the participants are undecided. A sequence of choice tasks is often used to narrow down upper and lower bounds to allow the researchers to compute the indifference point \cite{hardisty_how_2011}. The matching task directly asks for that point.

The comprehensive review by Frederick identified an extreme range of results across studies: The authors identified that the discount rates reported by studies ranged from negative to $\infty $, with most results ranging from 0\% to 500\% in what the authors describe as ``spectacular disagreement''\cite[p.~389]{frederick_time_2002}. They suggest that this may be partly due to the measurement method employed, a suggestion confirmed by later studies\cite{hardisty_how_2011}. They further discuss the broad range of factors that can in principle be used to explain \textit{why} participants discount future outcomes and which conditions and factors influence discounting behaviors. These include the distinction between time preference and time discounting; the role of uncertainty and anticipation; a distinction between monetary discounting of rewards, the different utility of rewards in time, and the temporal discounting of utility; several anchoring and priming effects; the framing of outcomes in terms of losses or gains; and the viscerality of outcomes, i.e., the degree of how vividly participants can imagine the outcome. For example, in the absence of precise uncertainty, participants will perceive future outcomes as less certain and apply additional discounting to the promised outcomes. This can partially be prevented by supplying quantitative uncertainty information, something we have adopted in our study. Hardisty et al. further discuss that the order in which choices are presented has a significant influence on the outcomes of the experiments \cite{hardisty_how_2011}. Weber et al. discuss psychological distance and viscerality as central factors that influence the degree of temporal discounting that can be observed~\cite{weber_experience-based_2006}, and Soman et al. explore how the discounting of outcomes that contain both gains and losses can be explained through the difference between the discounting losses and the discounting of gains \cite{soman_psychology_2005}. 

Overall, these studies and meta-studies provide a range of models and empirical guidelines that supported our study design and initial exploration of factors. We will return to these factors in the Discussion, but first focus on establishing what kind of behavior we can observe, and how we designed our study to establish an empirical baseline for examining temporal discounting in TD management.

\subsection{Discount rate elicitation}
\label{sec:discount_rate_elicitation}

Several methods have been developed to explore temporal discounting behaviors and elicit discount rates. The most important research design choices involve the presentation of the task or question that participants are asked to respond to; the variables that are manipulated within subjects and across subjects; and the calculation of implied discount rates from observed behaviors. Additionally, some studies explore the causal factors of temporal discounting or the effect of different research designs. Table \ref{tab:experiments} summarizes selected typical study designs that illustrate the range of choices. It compares a small range of the tasks, manipulated variables, sample populations and observed effects that are reported in the vast literature on temporal discounting.

As Hardisty et al.~\cite{hardisty_how_2011} demonstrate, the presentation of tasks influences the effect sizes obtained. For example, matching tasks exhibited lower discount rates than choice tasks~\cite{hardisty_how_2011,ahlbrecht_empirical_1997}. More generally, the range of choices presented to participants influenced the range of temporal discounting they exhibit~\cite{hardisty_how_2011}. For example, presenting them with options that correspond to extremely high discount rates elicits higher discount rates. The choice of which elicitation method to use should be based on a range of considerations including simplicity, congruence with the phenomenon under investigation, and ease of use for the participants~\cite{hardisty_how_2011}.

Traditionally, the literature on intertemporal choice has focused on a simple normative model of \textit{Discounted Utility} (DU) developed by Samuelson \cite{samuelson_note_1937}. In this model, the process of discounting is modeled as exponential: time-consistent and with a constant discount rate. The exponential model can be compared to the interest rate on loans and investments, which a ``rational'' decision-maker might supposedly use as a reference.

Because of pervasive inconsistencies with empirically observable behavior, it is often considered inadequate today and replaced by other models, e.g.,  \textit{hyperbolic discounting}. Each model supplies a different way to calculate discount factors.

%Discounted Utility concerns the desirability of some future event. People have a tendency to give greater value to rewards that are closer in time, and smaller value to rewards that are further into the future. People are said to discount the future reward. Consequently, they are willing to accept a smaller reward if they receive it sooner. To accept a delay in receiving the reward, it must thus be larger. The length of delay increases the factor by which the reward is discounted. There is extensive research examining how this factor changes over time under different conditions.

Studies that examine temporal discounting often ask participants to state under which conditions they would be indifferent between two temporally separated options, or when they would prefer one option over the other. For example, one study asked participants to state what amount of money would be equally attractive as receiving a smaller sum immediately \cite{thaler_empirical_1981}. Participants who were offered \$15 immediately stated that they would be indifferent to receiving \$30 in three months, \$60 in one year, or \$100 in three years (the values are medians of all responses). The annual discount rates for the three time periods in this example are 277\%, 139\%, and 63\%, illustrating that the change in rate is not constant -- participants would accept a smaller increase in reward if they had to wait longer.

All calculations are based on the stated amount that the person would require to prefer the future option (future value $F$), the amount available immediately (present value $P$), and the time in years between the two options $t$.  The annualized continuously compounded discount rate $DR_c$~\cite{samuelson_note_1937} relies on the future value $F$ calculated as Eq. (\ref{eq:FV}):

\begin{equation}
    F=P \times e^{DR_c \times t}
    \label{eq:FV}
\end{equation}

To obtain $DR_c$, the formula in Eq.~\ref{eq:DR} is used:

\begin{equation} 
    DR_c(F,P,t) = \frac {\ln{\frac{F}{P}}} {t}
    \label{eq:DR}
\end{equation}

% DF is not used in the paper
%From the discount rate $DR$, we can calculate the discount factor $DF$ as given in Eq.~\ref{eq:DF}:

%\begin{equation} 
%    DF(DR)= \frac {1} {1+DR}
%    \label{eq:DF}
%\end{equation}

The exponential model is relatively simple and follows the normative model of intertemporal choice. However, numerous studies have demonstrated deviations from the exponential model \cite{frederick_time_2002}. Everyday temporal discounting choices in most tasks are not based on a constant discount rate. \textit{Hyperbolic discounting}~\cite{mazur1987} is an alternative model that does not assume a constant discount rate \cite{doyle_survey_2012}. In this model, assigned value falls rapidly for earlier delays, but more slowly for longer delays, depending on the precise parameters used.

Different models yield different results. For example, some models amplify differences in discounting between small time intervals, while others are less sensitive to differences. Table~\ref{tab:model_comparison} shows discount rates for the data in the example above (by Thaler~\cite{thaler_empirical_1981}) using different models. The discount rates are not directly comparable between models, although the basic tendency in the data is captured by them all to different extents. Persons exhibit temporal discounting if their discount rates differ from zero. % In particular, the AUC approach yields values that are reversed compared to the other models (a larger value means less discounting in AUC).

The choice of model is a research topic in its own right. Several papers (e.g., \cite{samuelson_note_1937,mazur1987,myerson2001,hardisty_how_2011}) discuss the merits of different models, propose new models or variants of existing models, and examine their fit to different sets of empirical data. Ultimately, the intertemporal choice task should be chosen based on the real-life phenomenon of interest~\cite{hardisty_how_2011}, and the discount rate model must be chosen based on the data obtained while taking into account theoretical assumptions and comparability with related studies.

\begin{table}
    \centering
    \caption{Comparison of discount rates for different models with example data from~\cite{thaler_empirical_1981}}
    \label{tab:model_comparison}
    \begin{tabular}{|l|l|l|l|}
    \toprule
    Immediate reward & 3 months & 1 year & 3 years \\
    \midrule
    \$15 & \$30 & \$60 & \$100 \\
    \midrule\midrule
    Model & \multicolumn{3}{l|}{Discount rates} \\
    \midrule
    Exponential (Eq.~\ref{eq:DR}) & 3 & 1 & 1 \\
%    Hyperbolic (Mazur, Eq.~\ref{eq:hyperbolic})\textsuperscript{a} & 4 & 3 & 1.888 \\
    Hyperbolic~\cite{mazur1987} & 15 & 3 & 0.882 \\
%    Area under curve\textsuperscript{b} & 0.054 & 0.267 & 1 \\
    \bottomrule
    \multicolumn{4}{l}{\footnotesize Larger values represent more discounting.} \\
    \end{tabular}
    \vspace{-0.2cm}
\end{table}

\subsection{Summary}
Temporal discounting is a commonly studied phenomenon that arises in a range of situations and appears highly relevant for TD management. Prior research in intertemporal choice has focused on consumers and has gradually shifted from an initial focus on a constant discount rate to examining temporal shifts in preferences over time that are better represented by either hyperbolic discount curves or a close attention to subjective time perception. Many aspects deserve scrutiny, including differences in discounting gains, losses, and mixed outcomes; the design of the task used to elicit discount rates; and the interpretation of the findings.

In our study, we focus on establishing a baseline for further studies in software engineering by establishing whether there is a temporal discounting effect at all, and by discussing the shape of this effect in order to evoke opportunities for future research. Therefore, we investigate the following research questions:  

\renewcommand{\theenumi}{\textbf{RQ\arabic{enumi}}}
\begin{enumerate}
    \item Do software professionals discount uncertain future outcomes in decision making related to Technical Debt?  
    % \item Can we identify patterns of temporal discounting? 
    \item How does the temporal discounting behavior observed among software professionals compare to previous findings in intertemporal choice?
    % \item How can intertemporal choice research inform TD research?
    % \item Can theoretical models and findings from prior research on intertemporal choice shed light on the factors contributing to temporal discounting in TD management?
\end{enumerate}
\renewcommand{\theenumi}{\arabic{enumi}}

% interesting relationships between: 
%  consumer behavior <> marketing strategies vs. sw professionals <> SE management
%  personal behavior <> engineering methods

%\begin{comment}
%\begin{enumerate}
%    \item[RQ1] Do software professionals discount uncertain future outcomes? 
%    \item[RQ2] Can we identify patterns of temporal discounting? 
%    \item[RQ3] Is the temporal discounting behavior observed by software professionals similar to or distinct from that observed for consumers?
%    \item[RQ4] Can theoretical models and findings from prior research on intertemporal choice shed light on the factors contributing to temporal discounting in TD management?
%\end{enumerate}
%\end{comment}

\section{Research design and analysis}

\subsection{Questionnaire design}

To address our research questions, we developed an online questionnaire with an intertemporal choice task (see Fig.~\ref{fig:survey_task}). Participants were given a scenario with two options:
\begin{inparaenum}
    \item spend software project time earlier on implementing a planned feature (a short-term option); or
    \item integrate a software library with potential long-term benefit in terms of reduced maintenance effort. 
\end{inparaenum} We tested the effectiveness of the questionnaire in three pilots with experienced practitioners and academics known to us. Each iteration had new participants. After each iteration, we revised the questionnaire in terms of the scenario, the language used to frame the options and the overall design, based on the feedback received. The pilots also helped us to enhance our own understanding of the topic of investigation (i.e. intertemporal choices), and ways to analyze the data.

\begin{figure}
    \centering
    \fbox{\begin{minipage}{0.97\columnwidth}
    \centering
    \fbox{\begin{minipage}{0.97\columnwidth}
\footnotesize\sffamily
You are managing an \underline{N-years} project. You are ahead of schedule in the current iteration. You have to decide between two options on how to spend your upcoming week. Fill in the blank to indicate the least amount of time that would make you prefer Option 2 over Option 1.

\mbox{}

Option 1: Implement a feature that is in the project backlog, scheduled for the next iteration.
(five person days of effort).

\vspace{1mm}

Option 2: Integrate a new library (five person days effort) that adds no new functionality but has a 60\% chance of saving you \underline{\hspace{1em}} person days of effort over the duration of the project (with a 40\% chance that the library will not result in those savings).
\end{minipage}}

\vspace{1mm}

\begin{minipage}{0.99\columnwidth}
\footnotesize\sffamily
(The only difference here is the timeframe.)
\end{minipage}

\vspace{2mm}

\begin{minipage}{0.99\columnwidth}
\footnotesize\sffamily

For a project time frame of 1 year, what is the smallest number of days that would make you prefer Option 2?
\underline{\hspace{2ex}}
\vspace{2mm}

For a project time frame of 2 years, what is the smallest number of days that would make you prefer Option 2?
\underline{\hspace{2ex}}
\vspace{2mm}

For a project time frame of 3 years, what is the smallest number of days that would make you prefer Option 2?
\underline{\hspace{2ex}}
\vspace{2mm}

For a project time frame of 4 years, what is the smallest number of days that would make you prefer Option 2?
\underline{\hspace{2ex}}
\vspace{2mm}

For a project time frame of 5 years, what is the smallest number of days that would make you prefer Option 2?
\underline{\hspace{2ex}}
\vspace{2mm}

For a project time frame of 10 years, what is the smallest number of days that would make you prefer Option 2?
\underline{\hspace{2ex}}
\end{minipage}
\end{minipage}}
    
    \caption{Intertemporal choice task questionnaire (excerpt).}
\vspace{-0.3cm}
    \label{fig:survey_task}
\end{figure}

For the actual study, the participants were presented with a \textit{matching task}. They were asked to indicate the \textit{minimum} amount of potential time saving they would require to choose the long-term option over the short-term option. To avoid additional discounting due to absence of precise uncertainty in the future option, the scenario (option 2) specified that time-saving had a 60\% chance of being realized. This scenario was presented first as a 1-year project to establish a baseline preference (present value $P$)  without any priming from multiple time frames. In a subsequent stage, the question was repeated with a varying time frame as a 1, 2, 3, 4, 5, and 10-year project. The answer for the 1-year option from this subsequent stage was not used in the analysis. This allowed us to investigate whether and how participants would discount future choices in a TD scenario. The participants were also asked to answer a few demographic questions (e.g. gender, age, educational and professional qualification) in the same online questionnaire.

\subsection{Participants}
Participants were solicited from two large software development companies (Company 1 and Company 2) in Greece. The companies were chosen because both employ more than 100 employees, with a typical project (development) duration of 2 to 3 years and product lifetime of 10+ years. Moreover, both companies develop large-scale software systems (Company 1 develops enterprise systems for banks, government and educational institutes; Company 2 develops simulation software). Both use the Java programming language. 

\subsection{Data analysis}

We examined the survey results using basic statistical methods and calculated the discount rate using the exponential model with annualized continuous compounding (see Eq.~\ref{eq:DR}). Summary statistics were used to examine demographic data and describe the sample. Boxplots were used to obtain a visual overview of the temporal choice data. The median discount rate was plotted against the time horizon options in the scenario task to demonstrate the overall tendency. Individual discount rates were also plotted against the time horizon options. We examined the data for the two companies separately. 

The choice of discount rate model was based on the lack of available evidence in the field. Rather than testing the goodness of fit of various models and model parameters, we opted for the exponential model because it is commonly used in the intertemporal choice literature~\cite{hardisty_how_2011}, is easy to calculate and replicate, and is sufficient to determine whether discounting occurs or not. 

\begin{table}
\centering
\caption{Participant demographics}
\label{tab:participants}
\begin{tabular}{|p{2.8cm}|l|l|l|}
\toprule
& \textbf{Company 1} & \textbf{Company 2} & \textbf{Total} \\
\midrule
\textbf{No. of participants} & 23 & 10 & 33\textsuperscript{*} \\
\textbf{Female / Male / Other } & 5 / 18 / 0 & 0 / 10 / 0 & 5 / 28 / 0 \\
\textbf{Mean years of work experience (sd)} & 6.50 (5.66) & 7.70 (3.13) & 6.88 (4.98) \\ 
\bottomrule
\multicolumn{4}{l}{\textsuperscript{*}32 participants provided demographic data.}
\end{tabular}
\end{table}

\begin{figure*}
  \centering
  \subfloat[Responsibility]{\includegraphics[width=0.33\textwidth]{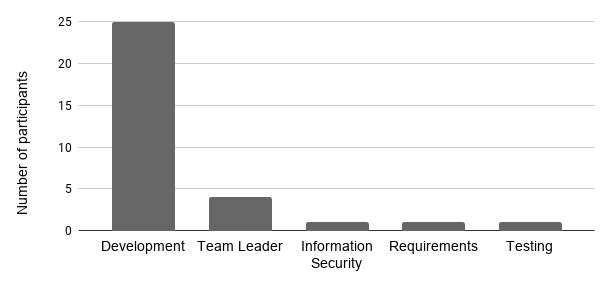}\label{roles}}
  \hfill
   \subfloat[Educational Qualification]{\includegraphics[width=0.33\textwidth]{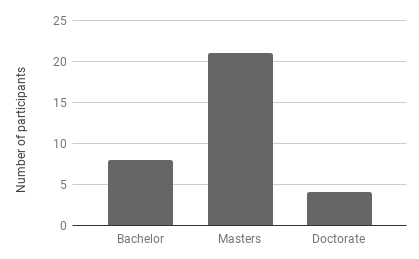}\label{education}}
  \hfill
  \subfloat[Work Experience]{\includegraphics[width=0.33\textwidth]{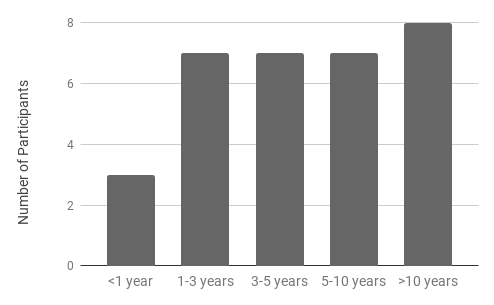}\label{workExperience}}
  \caption{Participant's demographics}\label{demographics}
  \vspace{-0.3cm}

\end{figure*}

\begin{figure*}
  \centering
  \subfloat[Company 1.]{\includegraphics[width=0.5\textwidth]{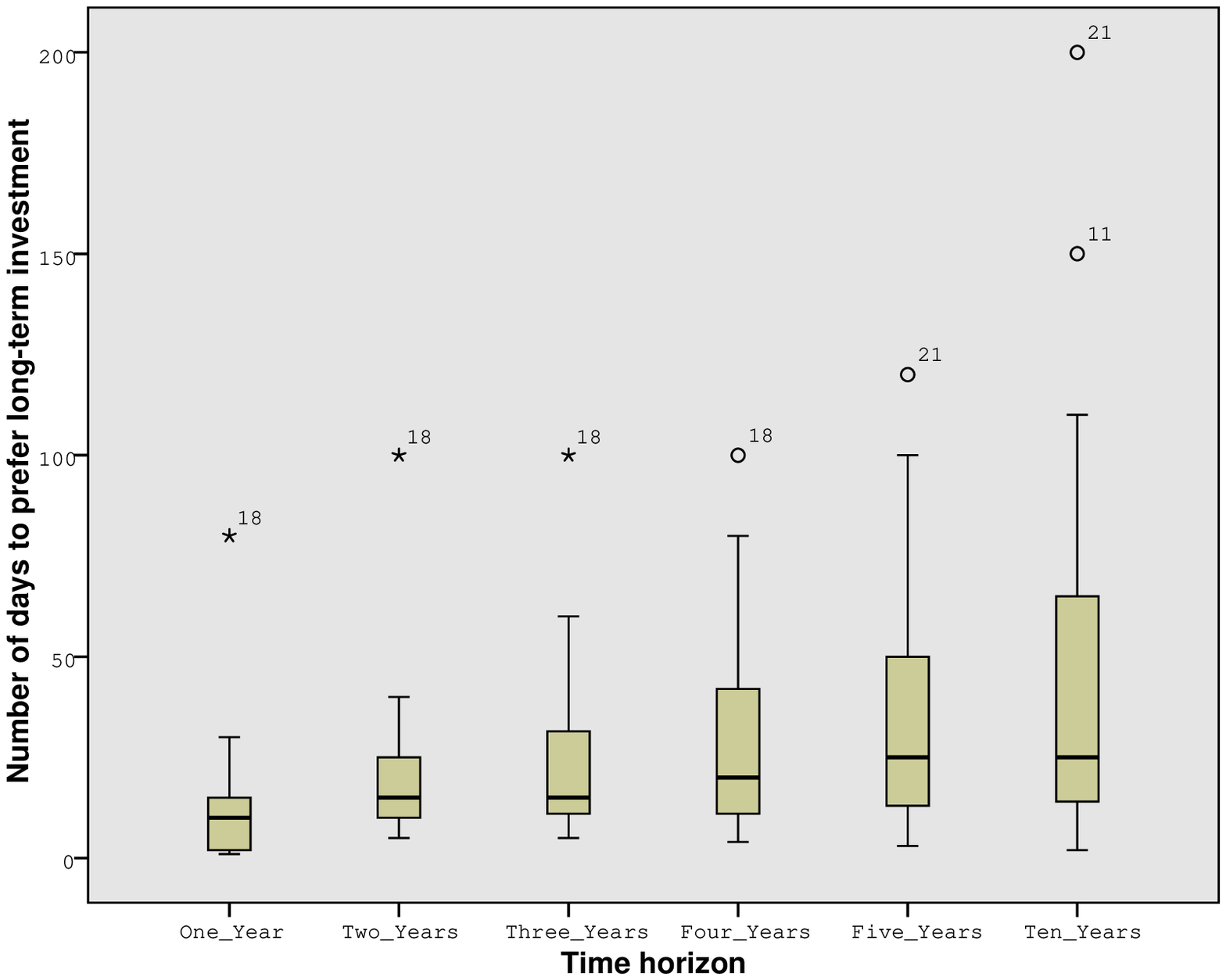}\label{boxplotcomp1}}
  \hfill
  \subfloat[Company 2.]{\includegraphics[width=0.5\textwidth]{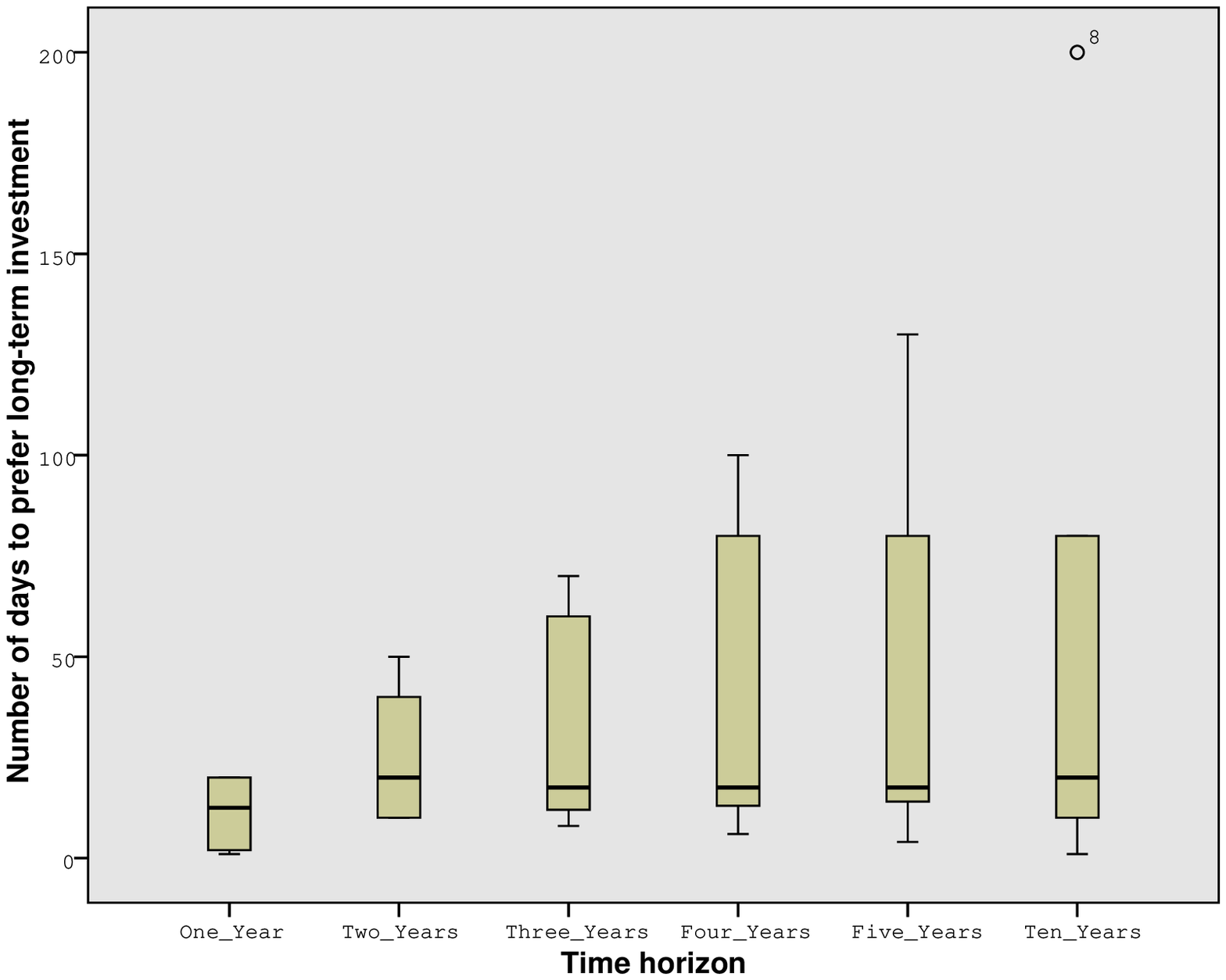}\label{boxplotcomp2}}
  \caption{Distribution of time savings required  to opt for long-term investment for varying project duration}
      \vspace{-0.3cm}
  \label{boxplot}
\end{figure*}

Similar to prior studies in intertemporal choice, we establish a present value for each participant through a singular choice, followed by a series of choices in which the time frame is manipulated to understand the effect of temporal delays on preference. The $60\%$ uncertainty was factored into the calculation of present and future values by deriving the expected value from the response provided by the participants -- for example, a response of 10 days results in an expected value of 6. The answer for the initial question about the 1-year project establishes the expected present value ($P$ in the equations in Section~\ref{sec:discount_rate_elicitation}). The answers for the subsequent questions about projects with 2 to 10 years time horizons then establish the expected future values ($F$ in the equations). 

\section{Results}

\subsection{Demographics}

%A convenience sample of 33 employees (23 from Company X; 10 from Company Y) participated.
%comprising of 5 females; 28 males) with a mean age of 25 years (standard deviation 6.07) participated.
After removing one duplicate entry, we obtained a sample of 33 employees. One participant provided no demographic details. We report demographics for the remaining 32 participants, but results on discount rates are drawn from all 33 participants. The mean age of participants was 34.3 years, with a standard deviation of 5.28. The individual responsibility in the company, the highest education qualification and work experience for all the participants are shown in Fig. \ref{demographics}. Table \ref{tab:participants} summarizes information on the number of participants, gender, as well as mean and standard deviation of work experience in both the companies. The data set with computed discount rates is available in a replication package \cite{dataset}.

\subsection{Do software professionals exhibit temporal discounting?}

If software professionals would value the future equally as the present, they would not exhibit temporal discounting. This would be expressed by discount rates around zero. We will describe results for each company and compare.

Fig. \ref{boxplot} illustrates the distribution of time savings required by the participants to opt for the long-term library investment for various project time durations. The box plots indicate the median number of days for each project duration (dark line), the 25th and 75th percentile (bottom and top of the boxes respectively), maximum and minimum values, as well as a few outliers. Note the significant spread of responses, which range from 1 day to 80 days. Numerically, responses below $8.3$ days imply an expected negative return on investment, which could be interpreted as risk-seeking behavior or as a choice made by considering external factors. 
%As discussed above, however, we do not compare future responses against the baseline option A, but against each participant's present value for option B.

For Company 1 (Fig.~\ref{boxplotcomp1}) it can be readily observed that for longer projects a gradually increasing number of days is required to prefer the long-term investment. This suggests the discounting of future choices in a technical debt scenario. The discount rates increase as the time horizon in which a benefit can be seen expands. The same holds for Company 2 (Fig.~\ref{boxplotcomp2}): the demand for higher savings by the long-term investment is also increasing as the pay-off spreads over a longer period, but the rate of increase is lower. Participants in both groups thus clearly exhibit temporal discounting.

\begin{figure*}
  \centering
  \subfloat[Company 1.]{\includegraphics[width=0.5\textwidth]{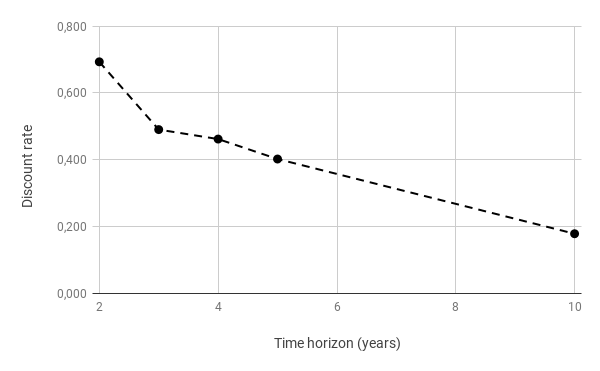}\label{discountratescomp1all.png}}
  \hfill
  \subfloat[Company 2.]{\includegraphics[width=0.5\textwidth]{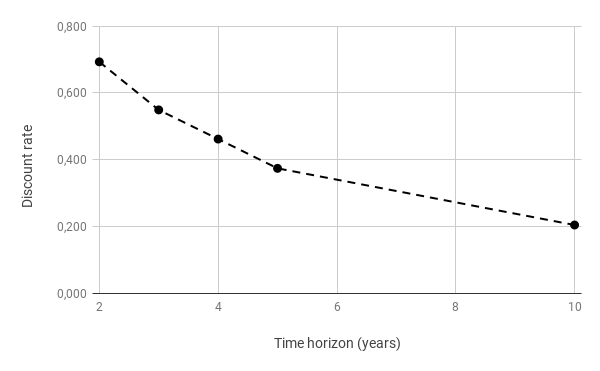}\label{discountratescomp2all.png}}
  \caption{Median discount rate as a function of time horizon}
\end{figure*}

% \begin{figure}
% 	\centering
% 	\includegraphics[width=1.0\linewidth]{figures/discountratescomp1.png}
% 	\caption{Median discount rate as a function of time horizon - responses with declining discount rate (Company 1).}
% 	\label{fig:discountratescomp1.png}
% \end{figure}

% \begin{figure}
% 	\centering
% 	\includegraphics[width=1.0\linewidth]{figures/chart.png}
% 	\caption{HYPERBOLIC DISCOUNTING Comp1 }
% 	\label{fig:chart.png}
% \end{figure}

%Just thinking that we in order to be clear here, we should state again that we are referring to annual DR (right?) and clarify which type of DR we are using - hyperbolic?
As discussed above, we compare future responses against each participant's present value for option 2 in order to establish the individual's discount rate per time horizon. For the majority of participants in the study, we observe that as the time horizon increases, the annual discount rate declines. The range of discount rates observed falls within the typical ranges observed by earlier studies on consumer behavior~\cite{loewenstein_time_2003}.

The median discount rate of the employees of Company 1 (see Fig.~\ref{discountratescomp1all.png}) reveals that as the time horizon increases, the annualized discount rate decreases. However, three participants exhibit no discounting at all, i.e. the amount of days required to prefer Option 2 remains constant throughout the time horizon. However, one participant specified preferences resulting in a negative expected return of investment. This implies they were either seeing other benefits (such as trying a new library) or maybe interpreting the time savings as a recurring benefit. 

% \textcolor{red}{From our discussion, that should be updated now - it's not a negative discount rate, but a keenness on avoiding TD :)}

\begin{figure*}
  \centering
  \subfloat[Company 1.]{\includegraphics[width=0.5\textwidth]{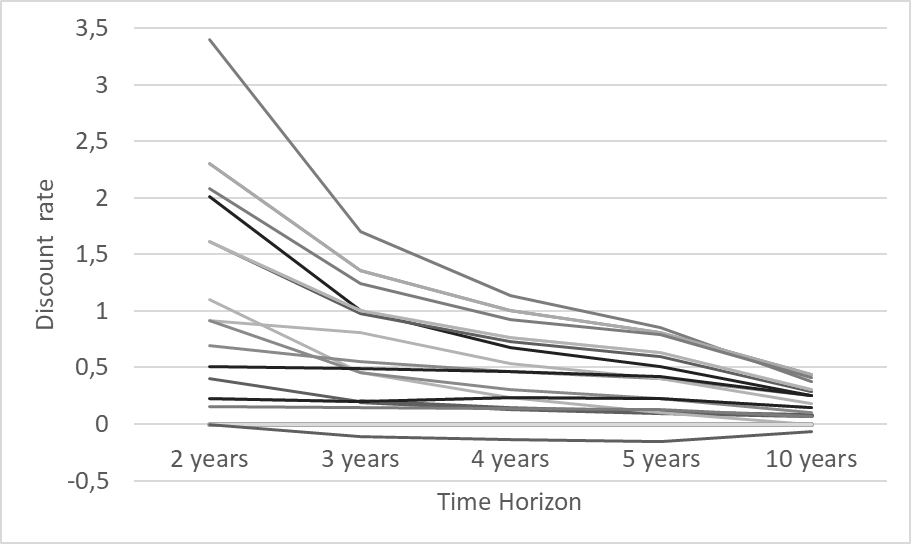}\label{fig:individualRatesComp1.png}}
  \hfill
  \subfloat[Company 2.]{\includegraphics[width=0.5\textwidth]{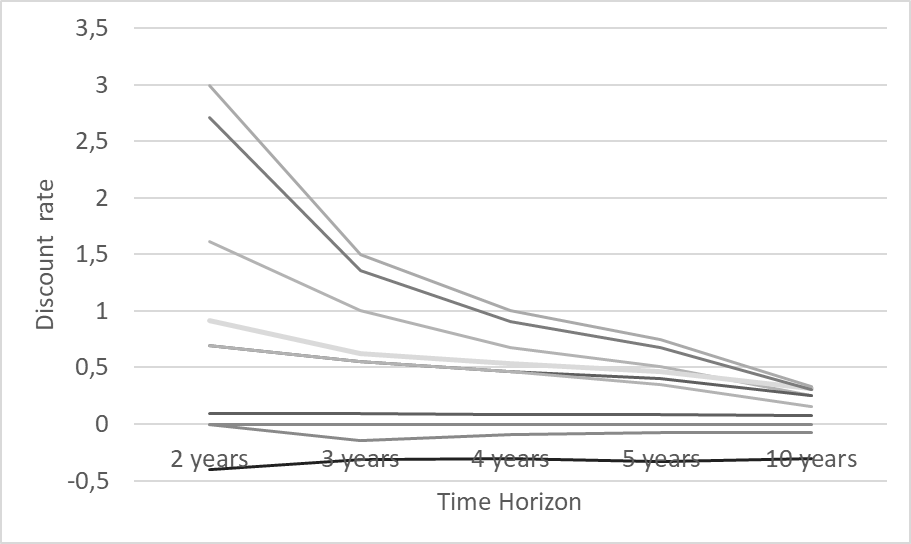}\label{fig:individualRatesComp2.png}}
  \caption{Individual discount rates as a function of time horizon}
  \vspace{-0.3cm}
\end{figure*}

For Company 2, the median discount rate evolution from all participants is shown in Fig.~\ref{discountratescomp2all.png}. The decreasing trend suggests that the discounting of future outcomes can be described by an exponential or even a hyperbolic curve.
An interesting finding of this study is that the discount curves for the two companies are almost identical. This almost perfect match between the curves possibly implies that developers of the same background, living in the same country and working at companies of the same (large) size, discount future in the same manner. This raises interesting question about how organizational and cultural factors influence temporal discounting.

%while the implied discount rates for Company 2 are  << the sentence had 'median' and 'implied' but as far as I can tell, it compares one and the same - simplified for readability. 
% Although the broad picture looks similar for both companies there are subtle differences. In particular, the median discount rate for the developers of both companies is exactly the same for time horizons of 2 and 4 years. For Company 2 is higher for the first three time horizons presented to them (1, 2 and 3 years), but lower for the longer time horizons (4, 5 and 10 years). In other words, the developers of Company 1 value a long-term investment on TD higher than the developers of Company 2, when the benefit is to be enjoyed within a relatively short period. Their preference appear to switch for longer time horizons. This could possibly be attributed to the fact that the project lifetime of Company 1 is somewhat shorter than that of Company 2 and thus project longevity is partially reflected on their temporal discounting. Another partial evidence for this claim is the fact that for Company 1 the declining trend for the discount rates over increasing time horizons seems to be more pronounced. However, further analysis would be required to systematically investigate the relation of environmental parameters (such as project longevity or project domain and complexity) to the reported discount rates with respect to TD related choices.

Median discount rates mask the preferences of individual developers and do not allow us to observe whether the declining trend is shared by all participants. Fig.~\ref{fig:individualRatesComp1.png} illustrates how the discount rates reported by each of the 23 developers in Company 1 shift according to the time horizon. It becomes evident that all developers exhibit a declining trend of temporal discounting as the time horizon of the project increases. Individual discount rates vary heavily for relatively short project duration. At one extreme, in order to prefer the long-term investment for a project of 2 years, a developer asked for savings which are 30 times higher (!) than the effort required to integrate the library in a project of 1 year. At another extreme, as already mentioned, one developer would opt for the integration of a library, even if this choice would yield lower savings in the future than the savings offered immediately. This implies that architectural TD mitigation is attractive for some people regardless of the associated effort to achieve it. However, it can be readily observed that choices converge when the time horizon increases: when the project duration reaches 10 years, almost all participants' choices imply a low discount rate. Fig.~\ref{fig:individualRatesComp2.png} illustrates the individual discount rates for Company 2. As was the case for the median discount rate, the individual participant's discount curves for the two Companies also exhibit remarkably similar patterns. One immediate question this raises relates to time perception: Is the difference in year 1 due to the difference in subjective time perception, or differences in opinion about the value of incurring vs.\ paying back Technical Debt? In other words, is it driven more by subjective time perception or by personal professional experience?

\input{tables/justification.tex}

\input{tables/additionalInfo.tex}

%which implies that opinions differ on how much benefit an investment on TD will bring within a year

\subsection{Which factors influence temporal discounting?}

The elicitation of individual discount rates establishes the phenomenon: Participants make choices \textit{as if} they were using discount rates. This does not establish the causal mechanisms that lead to the observed behaviors, of course. It is equally important to investigate the range of motives that drive developers' choices to eventually obtain a better understanding of intertemporal choice in the context of TD. This is ultimately a complex question that requires much more detailed research designs, but the study explored the multiple factors that participants considered explicitly through a set of follow-up questions, as is common in intertemporal choice studies.

Table \ref{tab:justification} lists representative responses to the question \textit{How did you make your choice?} It becomes evident that several developers appear to have followed some form of normative decision making approach for evaluating the expected utility of possible outcomes. This is evidenced by responses such as \textit{``I multiply the additional days with 0.6 ...''} or \textit{``the effort I would be pleased to save taking into consideration the 40\% failure''} and \textit{``I want to take back the (5-days) effort I invest plus 1 day as a marginal gain. So, 6 days is the minimum effort I want to gain. Since there is a 60\% probability for this to happen, in order to balance the risk and as an additional incentive I say 10 days''}. The responses reveal that many developers are well aware of the benefits of long-term investments and of the normative models of computing expected values.

The responses leave open, however, how the participants approached the intertemporal choice itself. For example, we cannot know whether the observed discount rate is the result of subjective time perception \cite{zauberman_discounting_2009}, time preference \cite{frederick_time_2002} or the differential discounting of outcomes involving gains and losses \cite{soman_psychology_2005}. Similarly, developers may be influenced by such aspects as whether this is their own project or someone else's \cite{amanatidis_developers_2018}.

Responses to the question \textit{What else would you need to know to make an informed choice?} shed light on parameters that participants consider, or would like to consider, in TD related decisions. Table \ref{tab:additional_info} lists representative responses. Team and project characteristics such as team size, skills, and remaining project time and resources are recurring concerns. This type of feedback is reasonable considering the inherent trade-off between assuring software quality and delivering in time.
%\vspace{-0.1cm}

\section{Discussion}

%\vspace{-0.1cm}

\subsection{Findings and Implications}

The results establish a first baseline for temporal discounting in Software Engineering by showing that software professionals exhibit temporal discounting: They value distant uncertain future outcomes lower than proximal ones at a rate proportional to the time horizon. The participants' responses show that their behaviors vary widely when shorter time horizons are considered and converge when time horizons expand.

Beyond this baseline of establishing that the effect can be observed, however, the study raises a plethora of new questions. Most importantly, the responses suggest that they consider a wide range of factors; use a range of different ways to think about this trade-off decision; and approach this logically and reasonably using their own experience and expertise. How exactly their reasoning leads them to make their decisions, however, is far from clear. 

Technical Debt (and other Software Engineering situations) provide very interesting temporal discounting scenarios because unlike consumer behaviors, they are taking place in a specialized domain of expertise in which specific normative models and approaches exist. As the study suggests, individuals still approach these situations using their common sense reasoning capabilities. How the normative methods that prescribe certain valuation behaviors interact with the perfectly reasonable common sense of individuals is one of the many open questions. Below, we discuss some of the areas we consider most fruitful for future investigations.

\subsection{Research Areas}

\subsubsection{The effects of temporal discounting in TD management}

If temporal discounting is as prevalent among software professionals as in the general population, this must be considered in the design of TD management methods and tools in at least two ways: First, as a source of behavior that causes TD, whether ``recklessly'' or ``inadvertently''~\cite{fowler_technical_2009}; second, as a factor that influences explicit choice in TD management decisions such as prioritization, scheduling or allocation. How  should  we  design intertemporal  choice situations  in TD and in SE considering these initial findings? 

Regardless of whether we consider this factor a bias or simply part of human nature, method design must consider temporal discounting appropriately, and this can only be done with some understanding of how these factors work. Intertemporal choice research in other domains, such as investment decision making, has explored highly relevant concepts for designing individual choices to improve delayed outcomes. TD management would do well to explore similar avenues, but we need a deeper understanding of causal factors first.

\subsubsection{Discount rate calculations}

In this initial study, we have opted for the most basic calculation of discount rates using the normative model of exponentially discounted utility. Many have pointed out, however, that this model is insufficient and does not address causal factors, varied behaviors, and shifting preferences. Future studies should explore hyperbolic models, subjective time perception, and other perspectives. 

\subsubsection{Subjective time perception in projects}

One area we consider very fruitful is to explore the subjective time perception of project duration, milestones and time horizons, and its role in temporal discounting. Prior research has provided strong arguments for the importance of time perception in temporal discounting~\cite{zauberman_discounting_2009}, and it is intuitively very reasonable to use relative perception, rather than mathematical discount rates, as a basis of explaining human choice. Prior research has also established the central role of framing in preferences \cite{tversky_framing_1981} and the importance of goal setting, incentives and time horizons for project management behavior \cite{abdel-hamid_impact_1999}. But how does the perception of project time frames create time horizons that frame the subjective perception of time?
    
\subsubsection{The discounting of mixed outcomes}

Since losses are discounted differently from gains, i.e., at a different rate over time -- outcomes that contain a mix of losses and gains can exhibit specific compound preference shifts. For example, Soman et al.~\cite{soman_psychology_2005} explored consumer preferences for a mixed outcome that involved a coupon that required some effort to redeem. Participant preferences could consistently be explained by discount rates that varied differently for the coupon reward and the effort to redeem the coupon. Specifically, it seemed that the effort was discounted more strongly than the reward, so that the prospect of redeeming a coupon in the future seemed more appealing than it did at a closer point in time. As a result, preference for the coupon became more pronounced in the distant future. Since efforts to incur or repay TD similarly involve a mix of gains and losses, it can be assumed that the perception similarly shifts over time: The closer in time the effort gets, the more commanding it appears.

\subsubsection{Social distance and viscerality}

In many situations involving temporal choices in TD, temporal discounting may interact with discounting based on social distance. For example, Amanatidis et al. showed that developers were more willing to recommend repayment of TD items when the code was in someone else's project~\cite{amanatidis_developers_2018}. This may interact in interesting ways with the discounting of mixed outcomes: After all, it's not the participants themselves that will have to refactor code or add documentation, but they can readily estimate the value this brings. It may simply be harder for them to empathize with the experience of having to invest the effort. This interaction of psychological distance and viscerality has been studied in other areas including environmentally related decision making and climate change~\cite{weber_experience-based_2006}. How does it influence TD?

\subsubsection{Individual, organizational and cultural factors}

It is very interesting to see that some participants did not exhibit any temporal discounting at all. Which factors influence these patterns of preferences? While two of them are the most experienced participants, one is the youngest, and no clear pattern can be distinguished in other regards either. Future studies should examine these patterns and explore what can be learned from those participants with long-term perspectives.

We speculate that organizational environments and broader cultural factors also have a marked influence on the trends that can be observed in intertemporal choice preferences. The participant population of the two companies in our study is homogeneous. We intend to replicate our study design across different companies and countries to begin comparing and identifying similarities and differences.
%^ difference between org1+2. Let's try to replicate and compare...
    %\item \textbf{Choice architecture design} 

\subsection{Threats to Validity}
The survey design presented here can be used to perform similar studies on intertemporal choice in software engineering. The identified patterns and the conclusions are of course subject to external validity threats that limit the ability to generalize these findings. The study focused on a limited factor of technical debt remediation, and closer inspection of other factors may reveal different results. The restriction of participants to two companies in one country necessarily reflect the beliefs and attitudes from a single culture and economy. Nevertheless, the results strongly suggest that many software professionals exhibit temporal discounting in their choices. Future work will explore how attitudes toward uncertainty and time perception may differ across populations.

%The design and interpretation of study results was defensive - our design does NOT maximize the effect...

The construct validity of any intertemporal choice study is threatened by the complexity of the presented scenario and the degree to which external effects and factors influence individual choices. For example, the temporal discounting of monetary rewards is strongly influenced by socio-economic background~\cite{harrison_estimating_2002}. In our study design, we followed good practice in intertemporal choice research by providing a fixed uncertainty indicator. In the follow-up questions of the study, only some of the subjects refer to the consideration of the 60\% probability of gaining a benefit in Option 2 in their decision making. (We did not ask them specifically whether they had considered it.) Some of the responses (e.g. \textit{\say{For bigger timeframes it is easier to take the risk}} and \textit{\say{...for too large projects, the risk (of investing on a library) is minimal}}, indicate that participants might even consider that for longer time horizons the investment on a library is safer. This is true in reality, but was meant to be factored out from the presented scenario by the given probability of savings. Some participants also considered additional causes of probability, disregarding that these were meant to be considered already in the 60\% probability. This threat is mitigated to a great extent by factoring out this probability in the discount rate calculation (i.e., we use the expected value by multiplying responses with $0.6$ for the calculations). Perhaps more importantly, some participants appear to have made choices based on an assumption that the savings are \textit{yearly}. As a result, their numbers may be biased too low. This is not a threat to the identified patterns, since temporal discounting is still present, but it might imply that temporal discounting in practice may be more pronounced than what is reported here.

\section{Conclusions}
The study's findings  establish that the intertemporal choice behaviors of developers are broadly consistent with what prior research has established for consumers. Just like consumers, developers exhibit temporal discounting and show significant differences in individual behavior. The findings establish the relevance of intertemporal choice theory and research for Software Engineering and raise a multitude of open research questions: Which factors affect temporal discounting? How and why does temporal discounting influence TD management decisions in particular and SE behavior more generally? And how should we design intertemporal choice situations in TD and in SE considering these findings? 

The paper establishes an empirical basis for understanding how software developers approach intertemporal choice and provides a model for conducting experiments about future discounting in TD management. As the first empirical study on intertemporal choice in Software Engineering, the paper provides a blueprint, establishes an empirical baseline and contributes essential empirically grounded insights to the theory of Technical Debt management.

%\section{Summary and Outlook}

\section*{Acknowledgements}
 Author contributions: CB suggested the initial idea. All developed and revised the research design and pilots. AC performed data collection and analysis. All participated in analysis, discussion and writing. Thanks to Dilip Soman for helpful early comments and suggestions. This research was partially supported by NSERC through RGPIN-2016-06640 and the KKS Foundation through the S.E.R.T. Research Profile at Blekinge Institute of Technology.
%\balance

\bibliographystyle{IEEEtran}
\bibliography{SMSBib}

\end{document}

%% file: tables/justification.tex
%!TEX root=conference_041818.tex

\begin{table*}
\rowcolors{2}{white}{gray!15}
\caption{Justification of choices (representative responses) }
\label{tab:justification}

\begin{center}
\begin{tabular}{|p{0.95\linewidth}|r}
% \hline
\toprule
\multicolumn{1}{|c|}{\textbf{How did you make your choice?}} 
\\\midrule
The library must give me back the days that we need to implement it plus the 5 days of the new feature time at least over every year\\
When taking a risk we expect some results that are worth the risk\\
The effort I would be pleased to save taking into consideration the risk of 40\% failure\\
I multiply the inserted days with 0.6 and i wanted the result to be enough for the library 5d + 5 days for the other task + 5 days to keep my upfront days
\\Library is always a tool you can reuse.. \\...(use of).. the new library .. might result in even more benefits in the future...
\\\bottomrule

\end{tabular}
\end{center}
\end{table*}

%% file: tables/additionalInfo.tex
%!TEX root=conference_041818.tex

\begin{table*}
\rowcolors{2}{white}{gray!15}
\caption{Additional information to make informed choice} 
\label{tab:additional_info}

\begin{center}
\begin{tabular}{|p{0.95\linewidth}|r}
% \hline
\toprule
\multicolumn{1}{|c|}{\textbf{What else would you need to know to make an informed choice?}} 
\\\midrule
The number of developers. The type of project (simple vs complex). The expected number of other dependencies.\\
The timeframe for my next task, as if the task is demanding I would prefer to start it earlier, instead of experimenting with an ambiguous library.\\
The knowledge level of the available resources. It differs when you assign a research task to a senior level person. It doesn't happen always but it is very possible a Senior Person to give better feedback in a week than 5 juniors. \\
Most probably the remaining time of the project
\\Total amount of people in the team
\\Are we at the beginning or the end of the project?
\\A) The specific time within the Project's timeline that the choice has to be taken. B) The overall person-days cost of the Project
\\\bottomrule

\end{tabular}
\vspace{-0.2cm}
\end{center}
\end{table*}

%% file: td.bbl
\begin{thebibliography}{10}
\providecommand{\url}[1]{#1}
\csname url@samestyle\endcsname
\providecommand{\newblock}{\relax}
\providecommand{\bibinfo}[2]{#2}
\providecommand{\BIBentrySTDinterwordspacing}{\spaceskip=0pt\relax}
\providecommand{\BIBentryALTinterwordstretchfactor}{4}
\providecommand{\BIBentryALTinterwordspacing}{\spaceskip=\fontdimen2\font plus
\BIBentryALTinterwordstretchfactor\fontdimen3\font minus
  \fontdimen4\font\relax}
\providecommand{\BIBforeignlanguage}[2]{{%
\expandafter\ifx\csname l@#1\endcsname\relax
\typeout{** WARNING: IEEEtran.bst: No hyphenation pattern has been}%
\typeout{** loaded for the language `#1'. Using the pattern for}%
\typeout{** the default language instead.}%
\else
\language=\csname l@#1\endcsname
\fi
#2}}
\providecommand{\BIBdecl}{\relax}
\BIBdecl

\bibitem{mcconnell_technical_2007}
\BIBentryALTinterwordspacing
S.~McConnell, ``Technical {Debt},'' 2007. [Online]. Available:
  \url{http://www.construx.com/10x_Software_Development/Technical_Debt/}
\BIBentrySTDinterwordspacing

\bibitem{cunningham_wycash_1992}
W.~Cunningham, ``The {WyCash} {Portfolio} {Management} {System},'' in
  \emph{Addendum to the {Proceedings} of {OOPSLA}}, ser. {OOPSLA} '92.\hskip
  1em plus 0.5em minus 0.4em\relax New York, NY, USA: ACM, 1992, pp. 29--30.

\bibitem{avgeriou_managing_2016}
\BIBentryALTinterwordspacing
P.~Avgeriou, P.~Kruchten, I.~Ozkaya, and C.~Seaman, ``Managing {Technical}
  {Debt} in {Software} {Engineering} ({Dagstuhl} {Seminar} 16162),''
  \emph{Dagstuhl Reports}, vol.~6, no.~4, pp. 110--138, 2016. [Online].
  Available: \url{http://drops.dagstuhl.de/opus/volltexte/2016/6693}
\BIBentrySTDinterwordspacing

\bibitem{becker_trade-off_2018}
C.~Becker, R.~Chitchyan, S.~Betz, and C.~McCord, ``Trade-off {Decisions}
  {Across} {Time} in {Technical} {Debt} {Management}: {A} {Systematic}
  {Literature} {Review},'' in \emph{Proceedings of {TechDebt} ’18:
  {International} {Conference} on {Technical} {Debt}, co-located with {ICSE}
  2018}.\hskip 1em plus 0.5em minus 0.4em\relax IEEE Press, 2018.

\bibitem{becker_intertemporal_2017}
\BIBentryALTinterwordspacing
C.~Becker, D.~Walker, and C.~McCord, ``Intertemporal {Choice}: {Decision}
  {Making} and {Time} in {Software} {Engineering},'' in \emph{Proceedings of
  the 10th {International} {Workshop} on {Cooperative} and {Human} {Aspects} of
  {Software} {Engineering}}.\hskip 1em plus 0.5em minus 0.4em\relax Piscataway,
  NJ, USA: IEEE Press, 2017, pp. 23--29. [Online]. Available:
  \url{https://doi.org/10.1109/CHASE.2017.6}
\BIBentrySTDinterwordspacing

\bibitem{frederick_time_2002}
\BIBentryALTinterwordspacing
S.~Frederick, G.~Loewenstein, and T.~O'donoghue, ``Time {Discounting} and
  {Time} {Preference}: {A} {Critical} {Review},'' \emph{Journal of Economic
  Literature}, pp. 351--401, 2002. [Online]. Available:
  \url{http://www.jstor.org/stable/2698382}
\BIBentrySTDinterwordspacing

\bibitem{soman_psychology_2005}
\BIBentryALTinterwordspacing
D.~Soman, G.~Ainslie, S.~Frederick, X.~Li, J.~Lynch, P.~Moreau, A.~Mitchell,
  D.~Read, A.~Sawyer, Y.~Trope, K.~Wertenbroch, and G.~Zauberman,
  ``\BIBforeignlanguage{en}{The {Psychology} of {Intertemporal} {Discounting}:
  {Why} are {Distant} {Events} {Valued} {Differently} from {Proximal}
  {Ones}?}'' \emph{\BIBforeignlanguage{en}{Marketing Letters}}, vol.~16, no.~3,
  pp. 347--360, Dec. 2005. [Online]. Available:
  \url{https://doi.org/10.1007/s11002-005-5897-x}
\BIBentrySTDinterwordspacing

\bibitem{loewenstein_time_2003}
G.~Loewenstein, D.~Read, and R.~F. Baumeister,
  \emph{\BIBforeignlanguage{en}{Time and {Decision}: {Economic} and
  {Psychological} {Perspectives} of {Intertemporal} {Choice}}}.\hskip 1em plus
  0.5em minus 0.4em\relax Russell Sage Foundation, Feb. 2003.

\bibitem{loewenstein_neuroeconomics_2007}
\BIBentryALTinterwordspacing
G.~Loewenstein, S.~Rick, and J.~D. Cohen,
  ``\BIBforeignlanguage{en}{Neuroeconomics},'' Dec. 2007. [Online]. Available:
  \url{http://www.annualreviews.org/doi/10.1146/annurev.psych.59.103006.093710#_i8}
\BIBentrySTDinterwordspacing

\bibitem{weber_experience-based_2006}
\BIBentryALTinterwordspacing
E.~U. Weber, ``Experience-{Based} and {Description}-{Based} {Perceptions} of
  {Long}-{Term} {Risk}: {Why} {Global} {Warming} does not {Scare} us ({Yet}),''
  \emph{Climatic Change}, vol.~77, no. 1-2, pp. 103--120, 2006. [Online].
  Available: \url{http://journals.scholarsportal.info/detailsundefined}
\BIBentrySTDinterwordspacing

\bibitem{thaler_empirical_1981}
R.~Thaler, ``Some empirical evidence on dynamic inconsistency,''
  \emph{Economics Letters}, vol.~8, no.~3, pp. 201--207, Jan. 1981.

\bibitem{coller_eliciting_1999}
\BIBentryALTinterwordspacing
M.~Coller and M.~B. Williams, ``\BIBforeignlanguage{en}{Eliciting individual
  discount rates},'' \emph{\BIBforeignlanguage{en}{Experimental Economics}},
  vol.~2, no.~2, pp. 107--127, Dec. 1999. [Online]. Available:
  \url{https://doi.org/10.1007/BF01673482}
\BIBentrySTDinterwordspacing

\bibitem{harrison_estimating_2002}
G.~W. Harrison, M.~I. Lau, and M.~B. Williams,
  ``\BIBforeignlanguage{en}{Estimating {Individual} {Discount} {Rates} in
  {Denmark}: {A} {Field} {Experiment}},''
  \emph{\BIBforeignlanguage{en}{American Economic Review}}, vol.~92, no.~5, pp.
  1606--1617, Dec. 2002.

\bibitem{zauberman_discounting_2009}
\BIBentryALTinterwordspacing
G.~Zauberman, B.~K. Kim, S.~A. Malkoc, and J.~R. Bettman,
  ``\BIBforeignlanguage{en}{Discounting {Time} and {Time} {Discounting}:
  {Subjective} {Time} {Perception} and {Intertemporal} {Preferences}},''
  \emph{\BIBforeignlanguage{en}{Journal of Marketing Research}}, vol.~46,
  no.~4, pp. 543--556, Aug. 2009. [Online]. Available:
  \url{https://doi.org/10.1509/jmkr.46.4.543}
\BIBentrySTDinterwordspacing

\bibitem{hardisty_how_2011}
\BIBentryALTinterwordspacing
D.~J. Hardisty, K.~Fox-Glassman, D.~Krantz, and E.~U. Weber,
  ``\BIBforeignlanguage{en}{How to {Measure} {Discount} {Rates}? {An}
  {Experimental} {Comparison} of {Three} {Methods}},'' SSRN, {SSRN} {Scholarly}
  {Paper} ID 1961367, Nov. 2011. [Online]. Available:
  \url{https://papers.ssrn.com/abstract=1961367}
\BIBentrySTDinterwordspacing

\bibitem{ahlbrecht_empirical_1997}
M.~Ahlbrecht and M.~Weber, ``An {Empirical} {Study} on {Intertemporal}
  {Decision} {Making} {Under} {Risk},'' \emph{Management Science}, vol.~43,
  no.~6, 1997.

\bibitem{samuelson_note_1937}
\BIBentryALTinterwordspacing
P.~A. Samuelson, ``A {Note} on {Measurement} of {Utility},'' \emph{The Review
  of Economic Studies}, vol.~4, no.~2, pp. 155--161, 1937. [Online]. Available:
  \url{https://www.jstor.org/stable/2967612}
\BIBentrySTDinterwordspacing

\bibitem{mazur1987}
J.~E. Mazur, ``An adjusting procedure for studying delayed reinforcement,'' in
  \emph{Quantitative analyses of behavior. The effect of delay and intervening
  events on reinforcement value}.\hskip 1em plus 0.5em minus 0.4em\relax
  Hillsdale, NJ: Erlbaum, 1987, vol.~5, pp. 55--73.

\bibitem{doyle_survey_2012}
\BIBentryALTinterwordspacing
J.~R. Doyle, ``\BIBforeignlanguage{en}{Survey of {Time} {Preference}, {Delay}
  {Discounting} {Models}},'' SSRN, {SSRN} {Scholarly} {Paper} ID 1685861, Apr.
  2012. [Online]. Available: \url{https://papers.ssrn.com/abstract=1685861.}
\BIBentrySTDinterwordspacing

\bibitem{myerson2001}
J.~Myerson, L.~Green, and M.~Warusawitharana, ``Area under the curve as a
  measure of discounting,'' \emph{Journal of the experimental analysis of
  behavior}, vol.~76, no.~2, pp. 235--243, 2001.

\bibitem{dataset}
\BIBentryALTinterwordspacing
C.~Becker, F.~Fagerholm, R.~Mohanani, and A.~Chatzigeorgiou, ``{Dataset and
  replication information for Temporal Discounting in Technical Debt: How do
  Software Practitioners Discount the Future?}'' Mar. 2019. [Online].
  Available: \url{https://doi.org/10.5281/zenodo.2595519}
\BIBentrySTDinterwordspacing

\bibitem{amanatidis_developers_2018}
\BIBentryALTinterwordspacing
T.~Amanatidis, N.~Mittas, A.~Chatzigeorgiou, A.~Ampatzoglou, and L.~Angelis,
  ``The {Developer}'s {Dilemma}: {Factors} {Affecting} the {Decision} to
  {Repay} {Code} {Debt},'' in \emph{Proceedings of the 2018 {Int.} {Conf.} on
  {Technical} {Debt}}.\hskip 1em plus 0.5em minus 0.4em\relax New York, NY,
  USA: ACM, 2018. [Online]. Available:
  \url{http://doi.acm.org/10.1145/3194164.3194174}
\BIBentrySTDinterwordspacing

\bibitem{fowler_technical_2009}
\BIBentryALTinterwordspacing
M.~Fowler, ``Technical {Debt} {Quadrant},'' 2009. [Online]. Available:
  \url{https://martinfowler.com/bliki/TechnicalDebtQuadrant.html}
\BIBentrySTDinterwordspacing

\bibitem{tversky_framing_1981}
A.~Tversky and D.~Kahneman, ``The {Framing} of {Decisions} and the {Psychology}
  of {Choice},'' \emph{Science}, vol. 211, no. 4481, pp. 453--458, 1981.

\bibitem{abdel-hamid_impact_1999}
\BIBentryALTinterwordspacing
T.~K. Abdel-Hamid, K.~Sengupta, and C.~Swett, ``The {Impact} of {Goals} on
  {Software} {Project} {Management}: {An} {Experimental} {Investigation},''
  \emph{MIS Q.}, vol.~23, no.~4, pp. 531--555, Dec. 1999. [Online]. Available:
  \url{http://dx.doi.org/10.2307/249488}
\BIBentrySTDinterwordspacing

\end{thebibliography}
